\documentclass[11pt]{article}

\usepackage{times}

\usepackage{epsfig}
\usepackage{amsmath}
\usepackage{dcolumn}% Align table columns on decimal point
\usepackage{bm}% bold math
\usepackage{graphicx}
\usepackage{wrapfig}
\usepackage{graphicx}
\usepackage{epstopdf}
\usepackage{epsfig}
\usepackage{calrsfs}
\usepackage{IEEEtrantools}
\usepackage{graphicx}
\usepackage{caption}
\usepackage{subcaption}

\newcommand{\nn}{\nonumber}

\newenvironment{sciabstract}{%
\begin{quote} \bf}
{\end{quote}}

\newcounter{lastnote}

\title{Nambu identity and collective modes \\ in superconductors and superfluid $^3$He}

\author
{Gavriil Shchedrin$^{\ast}$ and David M. Lee\\
\\
\normalsize{Department of Physics and Astronomy and Institute for Quantum Science and Engineering }\\
\normalsize{Texas A{\&}M University, College Station, TX 77843, USA}
\\\\
\normalsize{$^\ast$To whom correspondence should be addressed: shchedrin@mines.edu}
}

\date{}

\begin{document} 


\baselineskip24pt


\maketitle 
%\centerline{\today}


\begin{sciabstract}
Collective modes manifest themselves in a variety of different physical systems ranging from superconductors to superfluid  $^{3}$He. The collective modes are generated via the Higgs-Anderson mechanism that is based on the symmetry breaking double well potential. Recently collective modes were explored in superconducting  NbN and InO in the presence of a strong terahertz laser field.  In both cases a single collective mode that oscillates with twice the frequency of the superconducting energy gap $\Delta$ was discovered.
Superfluid $^{3}$He is the host for a whole variety of collective modes. In particular, in the superfluid $^{3}$He B-phase, two massive collective modes were found with masses $\sqrt{{8}/{5}}\Delta$ and $\sqrt{{12}/{5}}\Delta$. We show that for both cases of the superconducting films and for the superfluid $^{3}$He B-phase, the collective modes satisfy the Nambu identity that relates the masses of different collective modes to the energy gap parameter $\Delta$.
\end{sciabstract}

\begin{figure}[t!]
    \centering
    \begin{subfigure}[b]{0.3\textwidth}
        \includegraphics[angle=0, width=0.6\columnwidth]{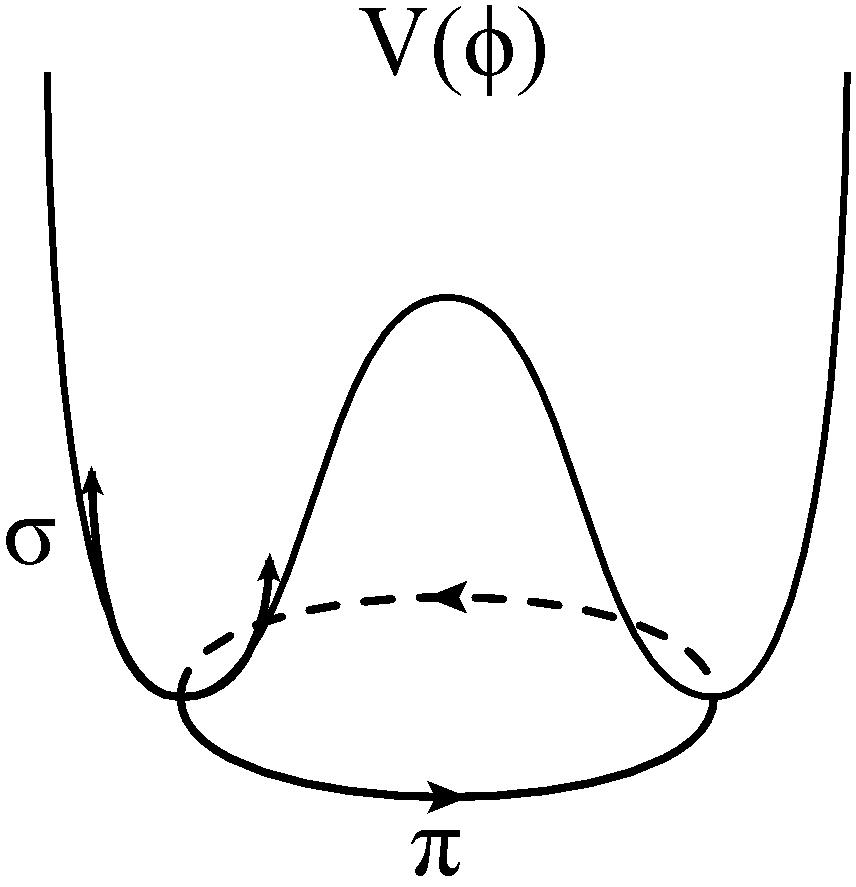}
        \caption{}\label{1a}
    \end{subfigure}
     \begin{subfigure}[b]{0.3\textwidth}
        \includegraphics[angle=0, width=0.8\columnwidth]{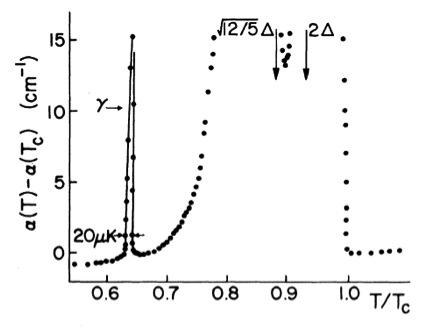}
        \caption{}\label{1b}
    \end{subfigure}\\
    \begin{subfigure}[b]{0.3\textwidth}
        \includegraphics[angle=0, width=0.6\columnwidth]{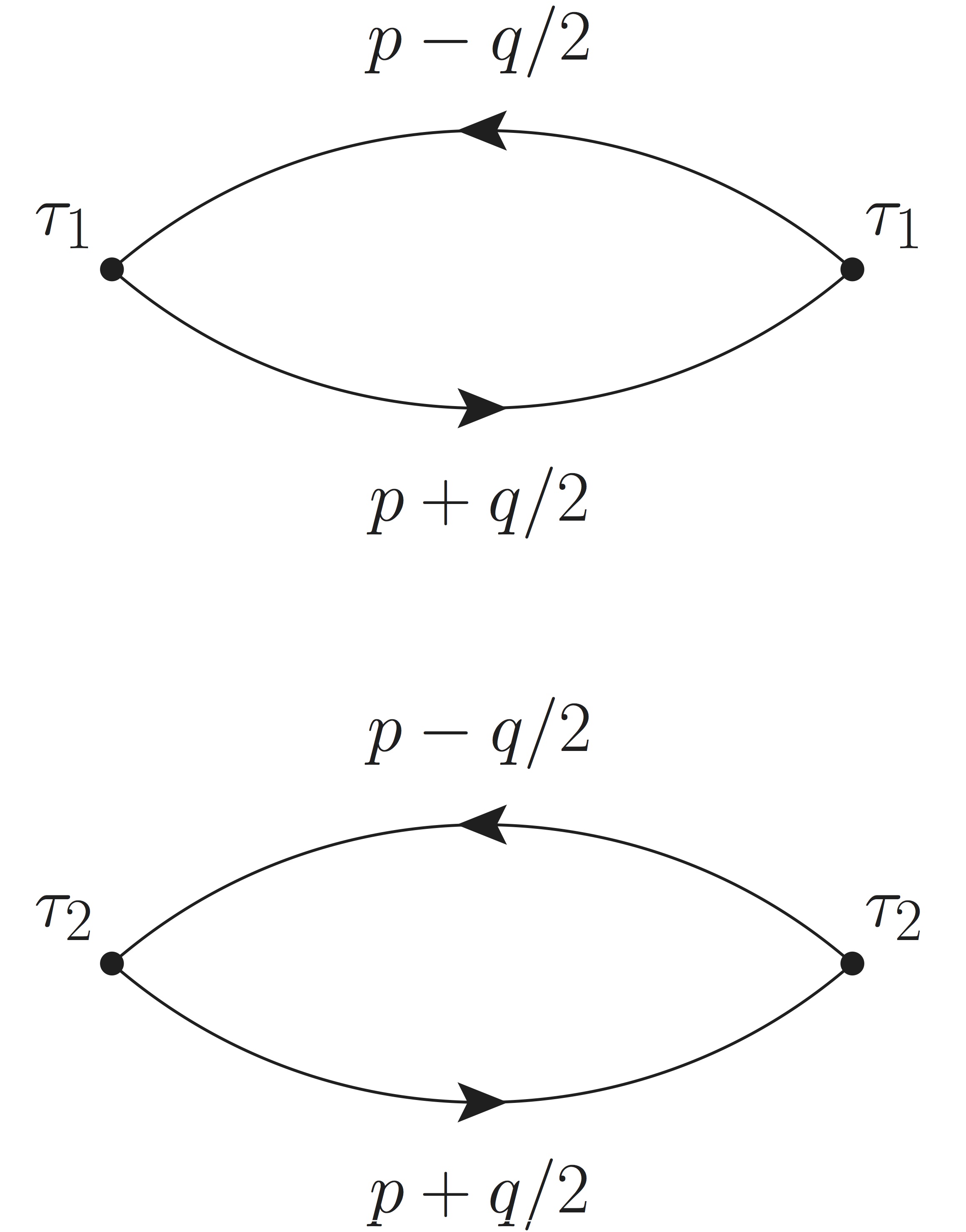}
        \caption{}\label{1c}
    \end{subfigure}
     \begin{subfigure}[b]{0.3\textwidth}
        \includegraphics[angle=0, width=0.85\columnwidth]{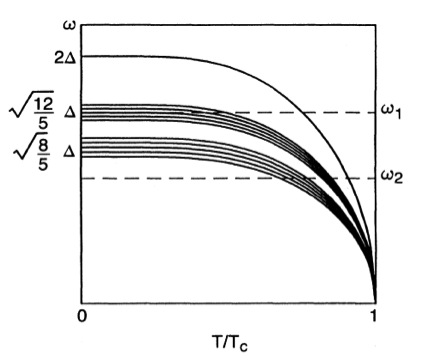}
        \caption{}\label{1d}
    \end{subfigure}
     \caption{(a) Double well potential with a polar massless mode $\pi$ and radial massive mode $\sigma$
 (b) Early evidence of the collective modes and pair breaking peaks in the superfluid $^3$He B-phase \cite{dmlee1, dmlee2, dmlee3, mast1, avenel1, calder1}
(c) Feynman diagram for the massive $\sigma$ mode and for the massless $\pi$ mode with 
the Pauli matrices $\tau_{1}$ and $\tau_{2}$ in the vertex, correspondingly. 
The Feynman diagram carries the four momentum $p$ along the loop, while the incoming momentum $q$ propagates through the chains of the Feynman loop diagrams and thus establishes the fermion-fermion interaction \cite{qft1,peskin1}.
(d) Schematic plot of the five-fold splitting of the collective modes in the superfluid $^3$He B-phase in a magnetic field \cite{dmlee1, dmlee2, dmlee3, mast1, avenel1, calder1}
  .}
\end{figure}

The recent discovery of the Higgs boson \cite{atlas1, cms1} triggered a great deal of excitement in the high energy physics and low temperature physics communities. In each of the fields of high energy and low temperature physics, the double well potential plays a central role in the field dynamics - it breaks the symmetry and creates collective modes \cite{qft1,peskin1} (see Fig.\ref{1a}). The collective modes in the Bardeen, Cooper, and Schrieffer theory of superconductivity \cite{bcs1} were predicted by Anderson \cite{anderson1} in 1958. Later Nambu \cite{nambu1,nambu2,nambu3,nambu4,lw1} discovered a simple and beautiful relation between the masses of different collective modes and the superconducting energy gap, known as the Nambu identity. For the case of up to two collective modes the Nambu identity reads,
\begin{eqnarray}\label{nambu1}
m_{1}^{2}+m_{2}^{2}=4\Delta^{2},
\end{eqnarray}
where $m_{1}$ and $m_{2}$ are the masses of the collective modes and  $\Delta$ is the energy gap parameter.

Recently, the collective modes were explored in superconducting NbN in the presence of a strong terahertz laser field. In their experiment, Matsunaga {\it et al} \cite{matsunaga1} used a broadband terahertz pulse to probe the dynamics of the superconducting energy gap. The result reveals that the energy gap oscillates at twice the terahertz driving frequency. Moreover the third harmonic oscillation of the induced superconducting current in the superconducting NbN was found. Thus Matsunaga {\it et al} discovered the collective mode $2\Delta$ in superconducting NbN. More recently Sherman {\it et al} \cite{sherman1} investigated the collective modes in superconducting NbN and InO via the terahertz spectroscopy, and 
along with Matsunaga {\it et al} observed the $2\Delta$ mode. It is worth mentioning that even earlier, Sooryakumar and Klein discovered the $2\Delta$ mode by the Raman scattering in superconductors  \cite{klein1}. 

We shall note that for a single massive collective mode mass in superconducting NbN, and another massless collective mode (i.e. $m_{2}={0}$), the Nambu identity (\ref{nambu1}) guarantees that the mass of the collective mode should be exactly $m_{1}=2\Delta$. 
In a recent communication, Anderson \cite{anderson2} has put the role of the Higgs field and the collective modes into the perspective of both condensed matter and high energy physics.

The collective modes were experimentally discovered previously in the superfluid $^3$He independently by Giannetta {\it et al} \cite{dmlee1,dmlee2}  (see Fig.{\ref{1b}}), and by Mast {\it et al} \cite{mast1} in 1980 and were eventually found to be in complete agreement with the Nambu identity. The collective modes found in the superfluid $^3$He B-phase have masses  $m_{1}=\sqrt{{8}/{5}}\Delta$ and $m_{2}=\sqrt{{12}/{5}}\Delta$. Thus the collective modes satisfy the Nambu relation (\ref{nambu1}) identically.

%\begin{figure}[h!]
%\begin{center}
%\includegraphics[angle=0, width=0.6\columnwidth]
%{figure10.eps}\\
%\end{center}
% \caption{(A) Double well potential with a polar massless mode $\pi$ and radial massive mode $\sigma$
% (B) Early evidence of the collective modes and pair breaking peaks in the superfluid $^3$He B-phase \cite{dmlee1, dmlee2, dmlee3, mast1, avenel1, calder1}
%(C) Feynman diagram for the massive $\sigma$ mode and the massless $\pi$ mode. Here $\tau_{1}$ and $\tau_{2}$ are the Pauli matrices.  The Feynman diagram carries the four momentum $p$ along the loop, while the incoming momentum $q$ propagates through the chains of the Feynman loop diagrams and thus establishes the fermion-fermion interaction \cite{qft1,peskin1}.
%(D) Schematic plot of the five-fold splitting of the collective modes in the superfluid $^3$He B-phase in a magnetic field \cite{dmlee1, dmlee2, dmlee3, mast1, avenel1, calder1}
%  .}
%\label{f1}
%\end{figure}
%%

\begin{figure}[t!]
\begin{center}
\includegraphics[angle=0, width=1.0\columnwidth]
{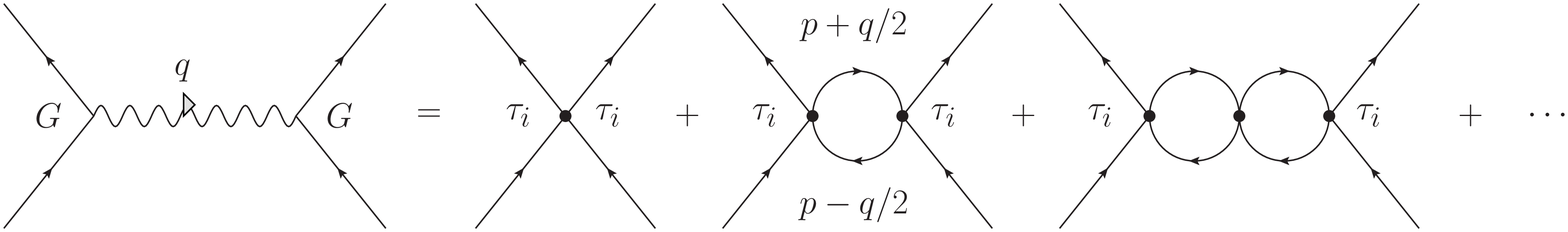}\\
\end{center}
 \caption{Meson exchange between the interacting fermions. The $\sigma(\pi)$-meson exchange is desciribed by the bubble diagram with the $\tau_{1} (\tau_{2})$ Pauli matrix in the
vertex.}
\label{f2}
\end{figure}

In this Comment we review the Nambu identity and compare it with the collective 
modes experimentally found in the superfluid $^{3}$He B-phase. The fermion-fermion interaction is mediated by the Feynman vacuum ``bubble'' diagrams depicted on Fig.{\ref{1c}} and Fig.{\ref{f2}}. The Feynman vacuum diagrams represent the meson exchange between the interacting fermions. The interaction carried by the massive $\sigma$ meson is described by the  vacuum Feynman  diagram,
\begin{eqnarray}\label{int1}
J_{\sigma}(q)=ig\int
{\frac{d^{4}p}{(2\pi)^{4}}}
{\rm Tr}
\left[
\tau_{1}
G\left(p+\frac{q}{2}\right)
\tau_{1}
G\left(p-\frac{q}{2}\right)
\right],
\end{eqnarray}
where we used the standard notation \cite{qft1,peskin1,nambu4}. 
The interaction carried by the massless $\pi$ meson is described by the  Feynman diagram,
\begin{eqnarray}\label{int2}
J_{\pi}(q)=ig\int
{\frac{d^{4}p}{(2\pi)^{4}}}
{\rm Tr}
\left[
\tau_{2}
G\left(p+\frac{q}{2}\right)
\tau_{2}
G\left(p-\frac{q}{2}\right)
\right].
\end{eqnarray}
Here $g$ is the coupling constant, $q$ is the incoming four-momentum, $p$ is the loop four-momentum, $\tau_{i}$ are the Pauli matrices, and $G\left(p\right)$ is the bare fermion propagator, 
\begin{eqnarray}
G(p)=\frac{p_{0}+\tau_{3}\varepsilon_{p}+\tau_{1}\Delta}{p_{0}^{2}-\varepsilon_{p}^{2}-\Delta^{2}+i\epsilon}.
\end{eqnarray}
The Feynman vacuum diagrams represent the meson exchange between the interacting fermions as one can clearly see from the Fig.(\ref{f2}).
Traces of the Pauli matrices can be directly obtained from,
\begin{IEEEeqnarray}{l}
{\rm Tr}
\left[
\tau_{i}\tau_{k}
\right]=2\delta_{ik},\\
{\rm Tr}
\left[
\tau_{i}\tau_{k}\tau_{l}
\right]=2i\varepsilon_{ikl},\\
{\rm Tr}
\left[
\tau_{i}\tau_{k}\tau_{l}\tau_{j}
\right]=2\left(\delta_{ik}\delta_{lj}+\delta_{ij}\delta_{kl}-\delta_{il}\delta_{kj}\right),
\end{IEEEeqnarray}
where $\varepsilon_{ikl}$ is the unit antisymmetric tensor of the third rank.

The integrals (\ref{int1},\ref{int2}) become
\begin{IEEEeqnarray}{l}\label{pimode}
J_{\pi}(q)=ig\int
{\frac{d^{4}p}{(2\pi)^{4}}}
{\rm Tr}
\left[
\tau_{2}
G\left(p+\frac{q}{2}\right)
\tau_{2}
G\left(p-\frac{q}{2}\right)
\right]=\\\nn
ig\int
{\frac{d^{4}p}{(2\pi)^{4}}}
\frac{2[p_{0}^{2}-q_{0}^{2}/4-\varepsilon_{p}^{2}-\Delta^{2}]}
{[(p_{0}-q_{0}/2)^{2}-\varepsilon_{p}^{2}-\Delta^{2}+i\epsilon]
[(p_{0}+q_{0}/2)^{2}-\varepsilon_{p}^{2}-\Delta^{2}+i\epsilon]},
\end{IEEEeqnarray}
and
\begin{IEEEeqnarray}{l}\label{sigmamode}
J_{\sigma}(q)=ig\int
{\frac{d^{4}p}{(2\pi)^{4}}}
{\rm Tr}
\left[
\tau_{1}
G\left(p+\frac{q}{2}\right)
\tau_{1}
G\left(p-\frac{q}{2}\right)
\right]=\\\nn
ig\int
{\frac{d^{4}p}{(2\pi)^{4}}}
\frac{2[p_{0}^{2}-q_{0}^{2}/4-\varepsilon_{p}^{2}+\Delta^{2}]}
{[(p_{0}-q_{0}/2)^{2}-\varepsilon_{p}^{2}-\Delta^{2}+i\epsilon]
[(p_{0}+q_{0}/2)^{2}-\varepsilon_{p}^{2}-\Delta^{2}+i\epsilon]}.
\end{IEEEeqnarray}
Performing the integration in the complex plane  \cite{qft1,peskin1,nambu4}  we obtain,
\begin{eqnarray}
J_{\pi}(q)=\frac{gN_{F}}{2}\int_{4m^{2}}^{\Lambda^{2}}
\frac{1}{k^{2}-q^{2}}
\left(
1-\frac{4m^{2}}{k^{2}}
\right)^{-1/2}
{dk^{2}},
\end{eqnarray}
\begin{eqnarray}
J_{\sigma}(q)=\frac{gN_{F}}{2}\int_{4m^{2}}^{\Lambda^{2}}
\frac{1}{k^{2}-q^{2}}
\left(
1-\frac{4m^{2}}{k^{2}}
\right)^{1/2}
{dk^{2}},
\end{eqnarray}
Here $\Lambda$ is the cut-off momentum, and $N_{F}$ is the density of states at the Fermi surface,
\begin{equation}
N_{F}=\frac{\nu p_{F}^{2}}{2\pi^{2}v_{F}}.
\end{equation}
It is defined in terms of the Fermi momentum $p_{F}$, the Fermi velocity, $v_{F}$, and the spin degeneracy $\nu=2$. If we introduce momentum densities for the $\pi$ and $\sigma$ modes, 
we find,
\begin{IEEEeqnarray}{l}
\rho_{\pi}(k^{2})\equiv{}
\left(
1-\frac{4m^{2}}{k^{2}}
\right)^{-1/2},
\\\nn
\rho_{\sigma}(k^{2})\equiv{}
\left(
1-\frac{4m^{2}}{k^{2}}
\right)^{1/2}.
\end{IEEEeqnarray}
Now we notice a simple relation 
\begin{eqnarray}
\rho_{\sigma}(k^{2})=\rho_{\pi}(k^{2})\left(
1-{4m^{2}}/{k^{2}}
\right)
\end{eqnarray}
This relation implies the connection between the massive $\sigma$ and massless $\pi$ modes,
\begin{eqnarray}
J_{\sigma}(q)=\frac{gN_{F}}{2}\int_{4m^{2}}^{\Lambda^{2}}
\frac{\rho_{\sigma}(k^{2})}{k^{2}-q^{2}}
{dk^{2}}=
\frac{gN_{F}}{2}\int_{4m^{2}}^{\Lambda^{2}}
\frac{\rho_{\pi}(k^{2})\left(
1-{4m^{2}}/{k^{2}}
\right)}{k^{2}-q^{2}}
{dk^{2}}.
\end{eqnarray}
This immediately leads to the $\Lambda$-independent, and therefore gauge invariant, expression,
\begin{equation}
J_{\sigma}(q)=
J_{\pi}(q)+\frac{4m^{2}}{q^{2}}
\frac{gN_{F}}{2}\int_{4m^{2}}^{\Lambda^{2}}
\left(
\frac{1}{k^{2}}-\frac{1}{k^{2}-q^{2}}
\right)\rho_{\pi}(k^{2}){dk^{2}}=
J_{\pi}(q)\left(1-\frac{4m^{2}}{q^{2}}\right)
+\frac{4m^{2}}{q^{2}}J_{\pi}(0),
\end{equation}
which can be written as
\begin{eqnarray}
\frac{q^{2}}{4m^{2}}J_{\sigma}(q)+\left(1-\frac{q^{2}}{4m^{2}}\right)J_{\pi}(q)=J_{\pi}(0).
\end{eqnarray}
With the self-consistent condition, $J_{\pi}(0)\equiv{1}$, and with $a\equiv{q^{2}/4m^{2}}$, we obtain \cite{nambu3}
\begin{eqnarray}\label{eq1}
aJ_{\sigma}(a)+(1-a)J_{\pi}(a)=1.
\end{eqnarray}
By adding and subtracting $a$, the identity (\ref{eq1})  can be rewritten as
\begin{eqnarray}\label{eq3}
-a(1-J_{\sigma}(a))=(1-a)(1-J_{\pi}(a)).
\end{eqnarray}
On the other hand, the summation of the geometric series generated by the Feynman bubble diagrams and depicted in Fig.\ref{f2} leads to 
\begin{eqnarray}
M_{\pi}=\frac{g}{1-J_{\pi}(q)},\\
M_{\sigma}=\frac{g}{1-J_{\sigma}(q)},
\end{eqnarray}
which is equivalent to the meson exchange with the masses $m_{\pi}$ and $m_{\sigma}$
correspondingly, i.e. 
\begin{eqnarray}
\frac{g}{1-J_{\pi}(q)}=-\frac{G^{2}}{q^{2}-m_{\pi}^{2}},\\\nn
\frac{g}{1-J_{\sigma}(q)}=-\frac{G^{2}}{q^{2}-m_{\sigma}^{2}}.
\end{eqnarray}
Together with $q^{2}=4\Delta^{2}a$ the identity $(\ref{eq3})$
transforms into,
\begin{eqnarray}
-a(4\Delta^{2}a-m_{\sigma}^{2})=(1-a)(4\Delta^{2}a-m_{\pi}^{2}),
\end{eqnarray}
which can be simplified as,
\begin{eqnarray}
am_{\sigma}^{2}+(1-a)m_{\pi}^{2}=4\Delta^{2}a.
\end{eqnarray}
We clearly see that at the bottom of the gap, that corresponds to $q_{0}=0$, i.e. for $a=0$, we obtain the massless $\pi$-meson, $m_{\pi}=0$, while at the top of the gap, $q_{0}=2\Delta$, i.e. for $a=1$, we obtain the single massive $\sigma$-meson with the mass $m_{\sigma}=2\Delta$.
Integrating over the gap,
\begin{eqnarray}
\int^{1}_{0}{da}
\left[am_{\sigma}^{2}+(1-a)m_{\pi}^{2}
\right]=\int^{1}_{0}{da}\;4\Delta^{2}a
\end{eqnarray}
we arrive at the Nambu mass identity \cite{nambu3},
\begin{eqnarray}
m_{\sigma}^{2}+m_{\pi}^{2}=4\Delta^{2}.
\end{eqnarray}

\newpage
For the total angular momentum $J=1$ the interaction mode becomes $\tau_{\alpha}[\sigma\times q]_{i}$. In terms of the loop integrals (\ref{pimode}, \ref{sigmamode}) we obtain
\begin{IEEEeqnarray}{l}
J^{(1)}_{\pi}(q)=
ig\int
{\frac{d^{4}p}{(2\pi)^{4}}}
\frac{2[p_{0}^{2}-q_{0}^{2}/4-\varepsilon_{p}^{2}+\Delta^{2}]}
{[(p_{0}-q_{0}/2)^{2}-\varepsilon_{p}^{2}-\Delta^{2}+i\epsilon]
[(p_{0}+q_{0}/2)^{2}-\varepsilon_{p}^{2}-\Delta^{2}+i\epsilon]}=J_{\sigma}(q),
\end{IEEEeqnarray}
\begin{IEEEeqnarray}{l}
J_{\sigma}^{(1)}(q)=
ig\int
{\frac{d^{4}p}{(2\pi)^{4}}}
\frac{2[p_{0}^{2}-q_{0}^{2}/4-\varepsilon_{p}^{2}-\Delta^{2}]}
{[(p_{0}-q_{0}/2)^{2}-\varepsilon_{p}^{2}-\Delta^{2}+i\epsilon]
[(p_{0}+q_{0}/2)^{2}-\varepsilon_{p}^{2}-\Delta^{2}+i\epsilon]}=J_{\pi}(q).
\end{IEEEeqnarray}
In other words the massless $\pi$-mode with a $J=0$ interaction mode becomes massive for the $J=1$ interaction mode, whereas the massive  $\sigma$-mode with $J=0$ interaction mode becomes massless for $J=1$.

The five-fold splitting (with $(2J+1)$ components) found in the superfluid $^3$He B-phase  is consistent with the $J=2$ modes (see Fig.\ref{1d}). Therefore we have to consider Feynman diagrams with the interaction mode that correspond to the total angular momentum $J=2$. 
For  $J=2$, the interaction mode becomes 
$\tau_{\alpha}[\sigma_{i}q_{k}-1/3\delta_{ik}(\sigma\cdot q)]$ and we obtain,
\begin{IEEEeqnarray}{l}
J^{(2)}_{\pi}(q)=
ig\int
{\frac{d^{4}p}{(2\pi)^{4}}}
\frac{2[p_{0}^{2}-q_{0}^{2}/4-\varepsilon_{p}^{2}+\Delta^{2}/5]}
{[(p_{0}-q_{0}/2)^{2}-\varepsilon_{p}^{2}-\Delta^{2}+i\epsilon]
[(p_{0}+q_{0}/2)^{2}-\varepsilon_{p}^{2}-\Delta^{2}+i\epsilon]}=\\\nn
\frac{3}{5}J_{\sigma}(q)+\frac{2}{5}J_{\pi}(q),
\end{IEEEeqnarray}
\begin{IEEEeqnarray}{l}
J_{\sigma}^{(2)}(q)=
ig\int
{\frac{d^{4}p}{(2\pi)^{4}}}
\frac{2[p_{0}^{2}-q_{0}^{2}/4-\varepsilon_{p}^{2}-\Delta^{2}/5]}
{[(p_{0}-q_{0}/2)^{2}-\varepsilon_{p}^{2}-\Delta^{2}+i\epsilon]
[(p_{0}+q_{0}/2)^{2}-\varepsilon_{p}^{2}-\Delta^{2}+i\epsilon]}=\\\nn
\frac{2}{5}J_{\sigma}(q)+\frac{3}{5}J_{\pi}(q),
\end{IEEEeqnarray}
where $J_{\pi}(q)$ and $J_{\sigma}(q)$ are the loop integrals (\ref{pimode}, \ref{sigmamode})
corresponding to the $J=0$ the interaction mode.

In other words the massless $\pi$-mode with the $J=0$ interaction mode becomes massive for the $J=2$ interaction mode with mass (see Figs.({\ref{1b}}, \ref{1d}))
\begin{IEEEeqnarray}{l}
m_{1}^{2}=\frac{3}{5}(2\Delta)^{2},\\\nn
m_{1}=\sqrt{\frac{12}{5}}\Delta,
\end{IEEEeqnarray}
whereas the massive   $\sigma$-mode with the $J=0$ interaction mode with mass $m_{\sigma}=2\Delta$ for the $J=2$ interaction mode becomes  (see Figs.({\ref{1b}}, \ref{1d}))
\begin{IEEEeqnarray}{l}
m_{2}^{2}=\frac{2}{5}(2\Delta)^{2},\\\nn
m_{2}=\sqrt{\frac{8}{5}}\Delta.
\end{IEEEeqnarray}

This relation means the that sum of squares of the boson masses on the left hand side (\ref{nambu1}) equals to the square of superconducting energy gap parameter. The important results obtained by Matsunaga {\it et al} \cite{matsunaga1} and independently by Sherman {\it et al}  \cite{sherman1} could stimulate an experimental search for collective modes inside the gap of non-s-wave superconductors.

We note that alternative derivations of the Nambu identity based on the kinetic equation for superfluid $^3$He were carried out independently by Sauls and Serene \cite{sauls1}, Nozieres \cite{noz1}, and Woelfle\cite{wolf1, wolf2,wolf3}. Moreover, Sauls and Serene \cite{sauls1} 
indicated corrections to the Nambu identity for the superfluid $^3$He-B phase. 
A possible indication of the $J=0$ gap mode in superfluid $^3$He-B phase was observed by  Peters and Eska \cite{eska1} who used high energy ultrasound pulses. 
The relation between masses of the composite Higgs bosons and the Nambu identity was recently discussed by Volovik and Zubkov \cite{volovik1}. 

The authors would like to thank Marlan Scully and  James Sauls for stimulating discussions.
The research was supported by the National Science Foundation Grants DMR-1209255,  PHY-1241032 (INSPIRE CREATIV), PHY-1068554, EEC-0540832 (MIRTHE ERC), and the Robert A. Welch Foundation Award A-1261.

\end{document}